\documentclass[paper=a4, fontsize=11pt]{article}
\usepackage{iftex,amsmath}
\iftutex
\usepackage{unicode-math}
\else
\usepackage{amssymb}
\def\z"{}
\def\UnicodeMathSymbol#1#2#3#4{%
 \ifnum#1>"A0
   \DeclareUnicodeCharacter{\z#1}{#2}%
  \fi}
\input{unicode-math-table}

\fi
\usepackage{amsthm}
\usepackage{enumitem}
\usepackage{cancel}
\usepackage{appendix}
\usepackage{graphicx}
\usepackage{geometry}
\usepackage{url}
\usepackage[colorlinks=true,linkcolor=blue]{hyperref}
\usepackage{sectsty}
\allsectionsfont{ \normalfont\scshape \textbf}
\newcommand{\mf}{\mathfrak}
\newcommand{\mc}{\mathcal}
\newcommand{\ad}{\mathfrak{a}^\dagger}
\newcommand{\as}{\mathfrak{a}}
\newcommand{\ep}{\varepsilon}
\newcommand{\da}{\downarrow}
\newcommand{\ua}{\uparrow}
\newcommand{\no}{\hat{N}}
\newlist{steps}{enumerate}{1}
\setlist[steps, 1]{label = Step \arabic*:}
\numberwithin{equation}{section}
\title{ The Case of Itinerant Magnetism in CaMn$_{2}$Al$_{10}$: Self-Consistent Renormalisation (SCR) Theory Study.}
\author{Bharathiganesh Devanarayanan $^{a,b}$$\,^{*}$, Akariti Sharma$^{a}$, Pratik D.Patel$^{a}$,\\ Navinder Singh$^{a}$\\
        \small $^{a}$Theoretical Physics Division, Physical Research Laboratory, Navrangpura Ahmedabad, India - 380009 \\
        \small $^{b}$Indian Institute of Technology, Gandhinagar, Palaj, Gujarat, India - 382355 \\\\
        \small $^{*}$Corresponding author:\,Bharathiganesh Devanarayanan; \tt{dbharathiganesh@gmail.com}
}
\date{} 

\begin{document}
\maketitle
\begin{abstract}
We have applied the powerful self-consistent renormalization theory of spin fluctuations for the system CaMn$_{2}$Al$_{10}$ discovered in $2015$ and was conjectured to be an itinerant magnet. We have calculated the inverse static {\it i.e.},  (paramagnetic) susceptibility and have compared it with the experimental data (Phys. Rev. B 92, 020413, 2015). The agreement is very good. We have calculated spin fluctuations at various temperatures and have also estimated the strength of the electronic correlation {\it i.e.},  ($I=0.3136$ eV) in the Hubbard Hamiltonian. Based on our quantitative explanation of the inverse static {\it i.e.},  (paramagnetic) susceptibility  data within the framework of SCR theory, we can decisively conclude CaMn$_{2}$Al$_{10}$ exhibits the phenomena of itinerant magnetism. Further, our DFT and DFT+U calculations corroborate the strong Mn-Al hybridization which is the key behind the itinerant magnetism in this system. Our estimated correlations strength will provide a foundation for further studies of itinerant magnetism in this system.    
 
\noindent   \end{abstract}

\section{Introduction}
Itinerant magnetism in Manganese(Mn) compounds is a rare phenomena. The reporting of a recent discovery of Mn-based weak itinerant magnetism in CaMn$_{2}$Al$_{10}$ compound is therefore quite interesting \cite{steinke2015camn}. In this compound Mn moments are very strongly hybridized with the conduction electrons and a much reduced moment results from correlations.
Generally, Mn compounds are insulators in which effective Coulomb interactions are sufficiently strong and examples include: Mn pnictides like
LaMnPO \cite{simonson2012antiferromagnetic,mcnally2014origin}, CaMn$_{2}$Sb$_{2}$ \cite{simonson2012magnetic}, and BaMn$_{2}$As$_{2}$ \cite{johnston2011magnetic}.  In metals, the Mn moments can be weakly hybridized, leading to the pronounced magnetic character of
systems like MnX (X = P,As,Sb,Bi) \cite{huber1964magnetic,heikes1955magnetic,ahlborn1975magnetic}, MnB \cite{lundquist1961saturation}, and
RMn$_{2}$X$_{2}$ (R = La,Lu,Y; X = Si,Ge) \cite{shigeoka1985magnetic,okada2002crystal}. There is another class of metallic compounds like MnSi \cite{steinke2015camn}, YMn$_{2}$ \cite{shiga}, and HfMnGa$_{2}$\cite{marques2011hffega}, in which the electronic fluctuations are so strong and there is no definite moment (or valence state).  Therefore, the current need is to investigate such Mn-based magnetic compounds on theoretical grounds in order to check their ability towards potential applications \cite{Aaina2021potentialapp}.
\par
According to the conventional wisdom if Mn-Mn distance in a Mn system is less than $2.7$ {\AA}, then the system exhibits the phenomenon of itinerant magnetism. But if the Mn-Mn  distance is more than $2.7$ {\AA} then that system must show localized magnetism. It is intuitively clear as less Mn-Mn distance leads to more electronic bandwidth. It turns out that the Mn-Mn distance in the system under study (CaMn$_{2}$Al$_{10}$) is greater than $2.7$ {\AA}. Thus it should exhibit the phenomenon of localized magnetism. However, this is in contradiction to what is observed in the experiment \cite{steinke2015camn}, and in our current theoretical study. The reason is strong Mn-Al hybridisation and it is explained in section \ref{sec4}. 

Experimentally, single crystals of CaMn$_{2}$Al$_{10}$ were grown as square rods as large as $1$x$1$x$10$ mm$^3$, where the crystallographic c-axis coincides with the rod axis. It is claimed that this compound could potentially fulfill the needs for low-dimensionality (with strong quantum fluctuations) and itinerant magnetism (with low ordering temperatures) \cite{pandey2020correlations}. The crystal structure of CaMn$_{2}$Al$_{10}$ had been determined from single crystal X-ray diffraction, and the composition had been verified by energy dispersive X-ray spectroscopy (EDS). 
\par
The magnetic properties of systems with itinerant electrons have been studied extensively and partly understood through the Hubbard model and the self-consistent renormalization (SCR) 
theory of spin fluctuations\cite{toru}. The SCR theory excellently captures the dynamics of spin fluctuations 
beyond the random phase approximation (RPA)\cite{izuyama1963band} by renormalizing the ground state including the effects of electronic correlations. Such a renormalization was first introduced by T. Moriya and A. Kawabata \cite{moriya1973effect} and has been studied by various authors  \cite{lonzarich1985effect,moriya1965ferro,moriya2006developments} for nearly ferromagnetic (FM) and paramagnetic (PM) metals. The SCR theory nicely takes in to account the influence of exchange-enhanced spin fluctuations on the thermodynamical quantities in a self-consistent manner. Gradually, this self-consistency has been proved essential for the theories of ferromagnetic \cite{lonzarich1985effect,moriya1973effect2} and anti-ferromagnetic \cite{hasegawa1974effect} materials. In this direction, a coupling theory of spin fluctuations in weakly ferromagnetic (FM) metals was developed later \cite{moriya2012}. The theory provides a mechanism, which explains the disagreement between the effective moment (obtained from the Curie constant) and the spontaneous moment of magnetic constituents \cite{toru}. This theory can quantitatively explain the Curie Weiss (CW) susceptibility for Ferromagnetic and Paramagnetic metals. 
Due to enhanced correlations in aforementioned systems ferromagnetic (FM) instabilities and \cite{stoner1938collective,slater1936ferromagnetism} and spin density wave (SDW) \cite{herring1951theory,herring1966magnetism} instabilities arise. At absolute zero ordering the spin fluctuations\cite{takahashi1986,takahashi2001,takahashi} with pronounced quantum character play an significant role in the measurement of quantities like magnetic susceptibility, specific heat, resistivity etc. Magnetic systems in this paramount edge can host a number fascinating phenomena like metal-to-insulator transition, non-Fermi-liquid to collective phases transition, superconductivity etc, and can be more clearly understood with the help of SCR theory \cite{moriya2006developments}.\\

In the next section we introduce very briefly the formalism of the SCR theory. In Section \ref{sec3} we discuss the results obtained from our SCR theory study. Section \ref{sec4} is devoted to DFT and DFT + U study, and we investigate the reasons behind the phenomenenon of itinerant magnetism in the system. In Section \ref{sec5} we conclude our results.

\section{Itinerant magnetism in CaMn$_{2}$Al$_{10}$: SCR theory studies}

Our starting point is the single band Hubbard Hamiltonian\cite{toru}:

\begin{equation}
    \mathcal{H} = \mathcal{H}_0 +\mathcal{H}'(\mathcal{I})
\end{equation}
\begin{equation}
    \mc{H}_{0} = \sum_{k, \sigma} \ep_{k} \ad_{k\sigma} \as_{k \sigma}
\end{equation}
\begin{equation}\label{8.3}
\mc{H}'(I) = I \sum_{k,k',q} \ad_{k+q\ua}\ad_{k'-q\da}\as_{k'\da}\as_{k\ua}
\end{equation}
The interaction part can be expressed in terms of the spin raising and lowering operators\cite{devanarayanan2021background}: 

\begin{equation}\label{8.4}
\mc{H}'(I) = \frac{1}{2} \mc{U} \no - \frac{1}{2} I \sum_{\mf{q}} \left\{ \mc{S}_+(\mf{q}) \mc{S}_-(\mf{-q}) \right\}
\end{equation}
here $\mc{N}$ is the total number of electrons, $\mc{U} = \mc{N}_{0} I$ is the intra-atomic exchange energy, $\mc{N}_{0}$ is the number of atoms in the crystal and the $\{\,\,\}$ anti-commutator. 
Next, the magnetic susceptibility in the unit of $\mu_{B}^2$ is given as
\begin{equation}\label{8.18}
\chi=\left[\left(\frac{\delta^2F(M)}{\delta M^2}\right)^{-1}\right]_{M=M^*}
\end{equation}
here, $M$ is the magnetization, $M^*$ is its saturation value and $F(M)$ is the total free energy as a function of $M$ {\it i.e.},
$M=N_{\downarrow}-N_{\uparrow}$ and
$N=N_{\downarrow}+N_{\uparrow}$.   
The partition function of the system in the presence of magnetic field is given as
\begin{equation}
Z(H)=Tr \left[e^{\frac{-[\mc{H}+ H M_{z}]}{K_{B}T}}\right]
\end{equation}
here $H$ is the magnetic field aligned along the $z$-axis, $\mc{H}$ is the Hamiltonian of the system and $M_{z}$ is the component of magnetization along $H$. Therefore, the free energy of the system is given as
\begin{equation}
F(H)=-T ln Z(H)    
\end{equation}
Further, the free energy can be expressed in terms of $M$ by using the Laplace transformation in the following way
\begin{equation}\label{8.19}
Z(M)=\frac{1}{2\pi i}\int_{-i\infty-\epsilon}^{i\infty+\epsilon} d\left(\frac{H}{T}\right) e^{\left(-\frac{H}{T}\right)M}Z(H)  \end{equation}
Define $F(M)=-T ln Z(M)$ is free energy for a given value of $M$ and we have
\begin{equation}\label{8.118}
\frac{F(M)}{T}=ln Z(M)\,\,\,\, ; \,\,\,  
Z(M)=e^{-\frac{F(M)}{T}}   
\end{equation}
Therefore from Eq. \eqref{8.19} and \eqref{8.118} we get
\begin{equation}
e^{-\frac{F(M)}{T}}=\frac{1}{2\pi i T}\int_{-i\infty-\epsilon}^{i\infty+\epsilon} d(H) e^{\left(-\frac{1}{T}\right) \left[(HM+F(H)\right]} 
\end{equation}
Using the saddle point approximation we get
\begin{equation}\label{8.24}
M=M^{*}=-\frac{\delta F(H)}{\delta H} \end{equation}
Where $M^{*}$ is the saturation value of $M$. From Eq. \eqref{8.24} (saddle point integral) we get
\begin{equation}
F(M^{*})=F(H)+HM^{*}    
\end{equation}
Physically, the free energy $F(M)$ can be estimated by calculating the free energy under an external magnetic field $H$ which gives rise to the magnetization $M$ and then subtracting the energy due to external field {\it i.e.}, $-HM^{*}$. Thus we express $F(M)$ as follows
\begin{equation}
  F(M) = F_{0}(M) + \int_{0}^{I}  \frac{dI}{I} \left< \mc{H}' (I)\right>_{M,I}
\end{equation}
Here, the total free energy of the magnetic system in terms of magnetization can be expressed as the sum of free energy terms due to free electrons, free energy contributed by the HFA (Mean-Field Contribution) and the term solely contributed by the correlation effects:

\begin{eqnarray}\label{free}
 F(M) = F_{0}(M) +\underbrace{F_{\textit{HF}}^{I}(M) + \varDelta F^{I}(M)}
\end{eqnarray}
\begin{center}
~~~~~~~~~~~~~~~~~~~~$F^{I}(M)$
\end{center}
In above equation $\varDelta F^{I}(M)$ is obtained as \cite{devanarayanan2021background}
\begin{equation}\label{8.245}
\Delta F^{I}(M)= -(1 / 2) \sum_{q} \int_{0}^{I} \mathrm{~d} I\left\{\left\langle\left[S_{+}(q), S_{-}(-q)\right]_{+}\right\rangle_{M, I}\right.\left.\\
-\left\langle\left[S_{+}(q), S_{-}(-q)\right]_{+}\right\rangle_{M, 0}\right\}
\end{equation}
Using the fluctuation-dissipation theorem\cite{toru} to express \eqref{8.245} in terms of the dynamical susceptibilities, we get 
\begin{equation}\label{8.244}
\Delta F^{I}(M)=-\frac{1}{2 \pi} \int_{-\infty}^{+ \infty} d \omega \operatorname{coth}\left(\frac{1}{2} \beta \omega\right) \operatorname{Im} \int_{0}^{I} d t \sum_{q}\left[\chi_{M,I}^{+-}(q,\omega)-\chi_{M, 0}^{+-}(q, \omega)\right]
\end{equation}
where $\chi_{M,I}^{+-}(q, \omega)$ is the transversal dynamic susceptibility under fixed values of $M$ and $I$:
\begin{equation}
\chi_{M,I}^{+-}(q, \omega)=i \int_{0}^{\infty} d t e^{i \omega t}\langle\left[S_{+}(q, t), S_{-}(-q)]\right\rangle_{M, I}
\end{equation}
After obtaining the renormalised free energy with the additional contributions from the electronic correlations, from \eqref{8.18} we obtain the susceptibility to be

\begin{equation}\label{sus1}
\chi=\frac{\chi_{0}}{1-\frac{1}{2} I \chi_{0}+\lambda(T)}
\end{equation}
where
\begin{equation}\label{sus2}
\lambda(T)=\frac{1}{2 \pi} \int_{-\infty}^{+\infty} d\omega \operatorname{coth}\left(\frac{1}{2} \beta \omega\right)G(\omega)
\end{equation}
and
\begin{equation}\label{8.4444}
G(\omega)=-\chi_{0} \operatorname{Im} \frac{\partial^{2}}{\partial M^{2}}\left(\int_{0}^{\infty} d I \sum_{q}\left[\chi_{M, I}^{+-}(q, \omega)-\chi_{M, 0}^{+-}(q, \omega)\right]\right).
\end{equation}
Since \eqref{free} is an exact expression for free energy, equation \eqref{sus1} along with equations \eqref{sus2} and \eqref{8.4444} is an exact expression for susceptibility. However \eqref{8.4444} requires the value of transversal dynamical susceptibilities as an input and they are not available apriori. One can then use the expression for transversal dynamical susceptibilities obtained using a random phase approximation in \eqref{8.4444}. The transversal dynamical susceptibilities under a fixed longitudinal molecular field $B$ by using a random phase approximation is obtained as\cite{toru}
 
\begin{equation}\label{rpasus}
\chi^{+-}(q, \omega)=\frac{\chi_{0}^{+-}(q, \omega)} {\left[1-I \chi_{0}^{+-}(q, \omega)\right]}\\
\end{equation}  
Substituting \eqref{rpasus} in \eqref{8.4444}, we get the following expression:
\begin{equation}\label{conrpa}
\begin{aligned}
G(\omega)=&-\xi \alpha^{2}(4 \pi)^{-1} \operatorname{Im} \int \mathrm{d} q\left\{f_{0}\left(\partial^{2} f_{0} / \partial B^{2}\right)\left(1-\alpha f_{0}\right)^{-1}\right.
\left.+\left(\partial f_{0} / \partial B\right)^{2}\left(1-\alpha f_{0}\right)^{-2}\right\}
\end{aligned}
\end{equation}
with
\begin{equation}
\left.\begin{array}{l}
f_{0}=f_{0}(q, \omega+\mathrm{is})=\chi_{0}^{+-}(q, \omega+\mathrm{is}) /\left(\chi_{0} / 2\right) \\\\
\xi=\left(k_{\mathrm{F}}^{3} / 2 \pi^{2} \varepsilon_{\mathrm{F}} \chi_{0}\right), \quad \alpha=I \chi_{0} / 2
\end{array}\right\}
\end{equation}
where $\zeta$ equals to $1$ for an electron gas model at $T=0\mathrm{~K}$ and $k_F$ ($\epsilon_F$) is the Fermi wave-vector (Fermi-energy). But as shown in \cite{devanarayanan2021background} for the static and long wavelength limit \textit{i.e.}, ($q\rightarrow 0$ and $\omega\rightarrow 0$), the RPA susceptibility is the same as the one obtained from Stoner theory. Thus one has to employ a modified random phase approximation that adjusts the RPA susceptibility at $q = \omega = 0$. This is realised by replacing $(1-\alpha) / \alpha$ in \eqref{conrpa} by
\begin{equation}\label{sc1}
\delta=\chi_{0} / \alpha \chi=2 / I \chi=(1-\alpha+\lambda) / \alpha.
\end{equation}
With this \eqref{conrpa} becomes:
\begin{equation}\label{sc2}
\begin{aligned}
G(\omega)=-\xi(4 \pi)^{-1} Im \int \mathrm{d} q\{\alpha f_{0}\left(\partial^{2} f_{0} / \partial B^{2}\right)\left(\delta+1-f_{0}\right)^{-1}+(\partial f_{0} / \partial B)^{2}(\delta+1-f_{0})^{-2}\}.
\end{aligned}
\end{equation}
One can thus obtain the magnetic susceptibility by solving the equations \eqref{sus2}, \eqref{sc1} and \eqref{sc2} self consistently. This method of modified random phase approximation has also been used for the calculation of transversal dynamical susceptibilities and Curie temperature for  heavy fermion like materials \cite{coleman2001fermi,gegenwart2008quantum}.

\section{Results and discussions}\label{sec3}
\begin{figure}[h]
    \centering
     \includegraphics[scale = 0.25]{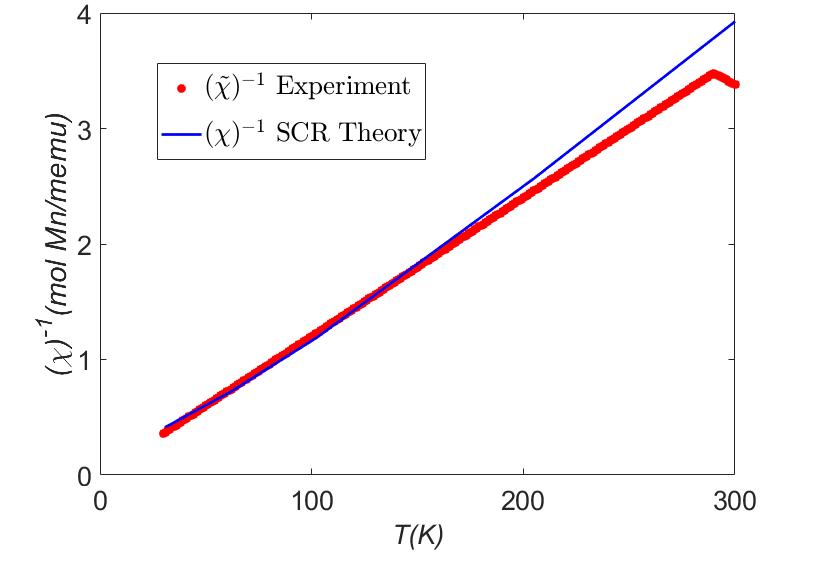}
    \caption{$(\chi)^{-1}$ vs $T$ plots at $q_{c}=2.5$ and $\alpha=6.205$ in the SCR theory along with the experimental data \cite{steinke2015camn}.}
    \label{fig1}
\end{figure}

We have performed numerical calculations to obtain the static susceptibility by solving the equations \eqref{sus2}, \eqref{sc1} and \eqref{sc2} self consistently and have compared it with the experimental data available\cite{steinke2015camn}. The susceptibility obtained through SCR theory is in dimensionless form \textit{i.e.} $\chi_{0}/\chi$ as a function of $T/T_{F}$. Hence to compare the theoretical results with that of experiments we will need the values of $\chi_{0}$ and $T_{F}$. The value of $\chi_{0}$ is reported to be $3.2\times10^{-5}$ emu/ mol Mn\cite{steinke2015camn}. However the value of $T_{F}$ has not been reported in the literature. We have calculated $T_{F}$ using the value of the Sommerfield coefficient from the specific heat data reported in \cite{steinke2015camn} and it is estimated to be $\simeq 1025K$. With this we perform numerical calculations for various values of the interaction parameter $\alpha$ and the cut-off parameter $q_{c}$. We obtain a best fit for the experimental data for $\alpha = 6.205$ and $q_{c} = 2.5$ as shown in figure \ref{fig1}.  

\begin{figure}[h]
\centering
\includegraphics[scale = 0.25]{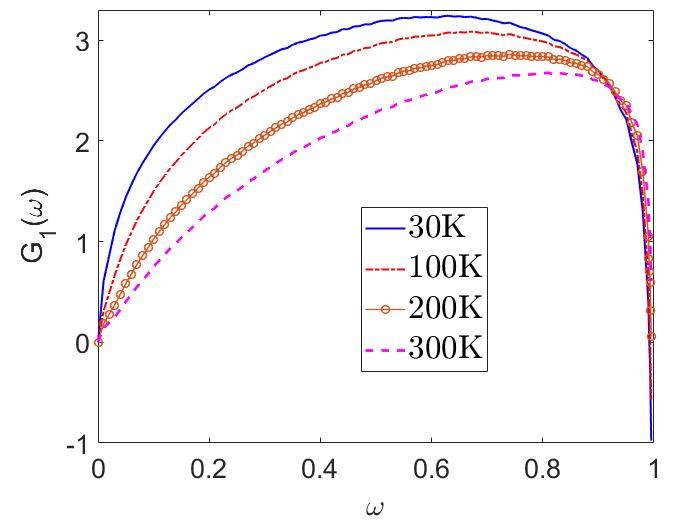}
\includegraphics[scale = 0.25]{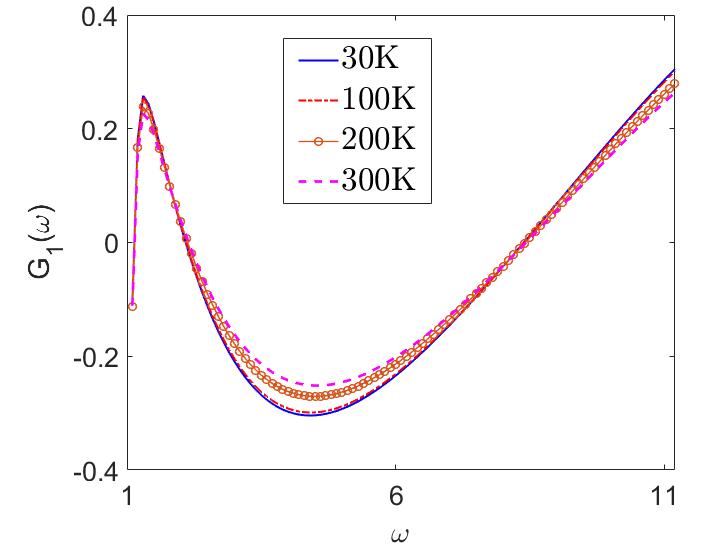}
\caption{$G_{1}(\omega)$ for $\omega<1$ in [panel (a)] and for $\omega>1$ in [panel(b)] at indicated $T$-values for $q_c=2.5$.}
    \label{fig2}
\end{figure}

\begin{figure}[h]
    \centering
    \includegraphics[scale = 0.25]{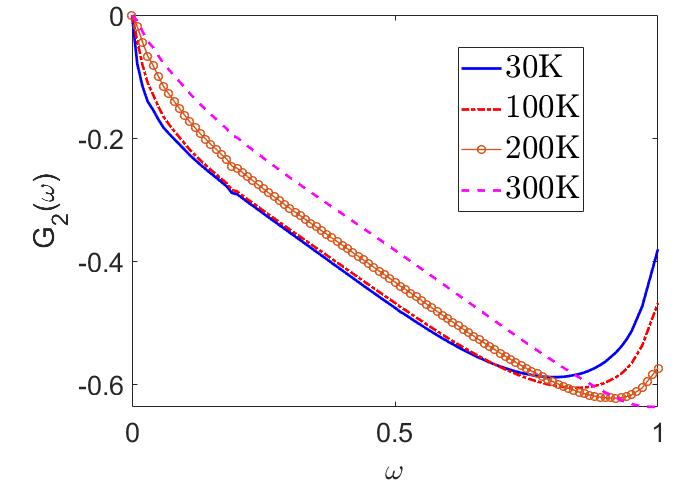}
    \includegraphics[scale = 0.25]{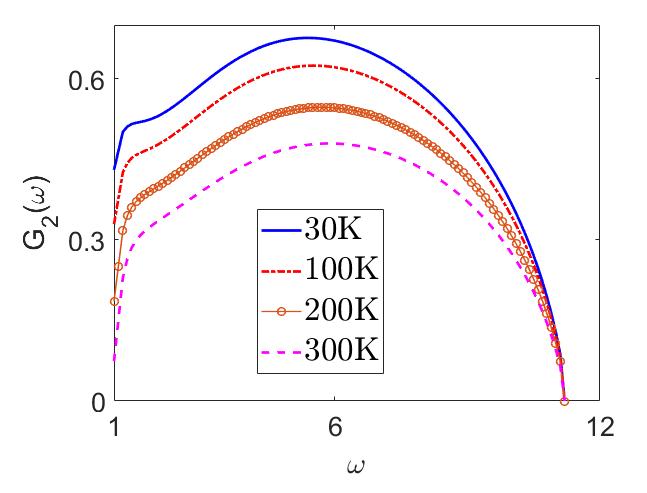}
    \caption{$G_{2}(\omega)$ for $\omega<1$ in [panel (a)] and for $\omega>1$ in [panel(b)] at indicated $T$-values for $q_c=2.5$.}
    \label{fig3}
\end{figure}

One can see that there is a very good agreement between the results obtained from SCR theory and the experiments for temperatures below $150 K$ and this is to be expected as SCR theory has been known to give accurate results for lower temperatures with respect to the RPA. With the optimal interaction parameter $\alpha$ that has been obtained, we have calculated the interaction coefficient in the Hubbard Hamiltonian to be $I = 0.3136$ eV. In the process of obtaining the inverse susceptibility we have also calculated the spin fluctuations $G(\omega)( = \alpha\, G_{1}(\omega) + G_{2}(\omega))$ for all temperatures. The spin fluctuations for some selected temperature values are shown in figures \ref{fig2} and \ref{fig3}. 

By looking at figures \ref{fig2} and \ref{fig3} we can make the following observations: (a) The amplitude of spin fluctuations is large for lower temperatures and less for higher temperatures indicating that the amplitude of spin fluctuations decrease with increase in temperature. This is reasonable and indicates that the effect of electronic correlations is pronounced at lower temperatures. (b) The frequency corresponding to the maxima of the spin fluctuations increases with increase in temperature.
\par
For clarity, $G(\omega)$ can be separated in two parts {\it i.e.}, $G_{1}(\omega)$ and $G_{2}(\omega)$ (for details see appendix). In figure \ref{fig2}(a), it is seen that $G_{1}(\omega)$ is positive and its magnitude decreases with increase in $T$. In this $\omega$ regime, the results of $G(\omega)$ can also be interpreted as: with increase in $T$ the $G_{1}(\omega)$ becomes less repulsive. On the other hand it clear from figure \ref{fig3}(a) that $G_{2}(\omega)$ is negative in low-$\omega$ regime and becomes less negative with increase in $T$. This behaviour of $G_{2}(\omega)$ is in complete contrast with $G_{1}(\omega)$ give rise to attractive correlations in low-$\omega$ regime. Notably, both $G_{1}(\omega)$ and $G_{2}(\omega)$ becomes singular around $\omega \simeq 1$ for very low-$T$ ($\simeq 0K$) and this singularity is chopped off with further increase in $T$. Next, in figure \ref{fig2} (b) it is found that $G_{1}(\omega)$ oscillates in $\omega>1 $ region {\it i.e.}, it is negative for $2\leq\omega\leq6$ and becomes positive for $\omega>6$. While as $G_{2}(\omega)$ remains positive in high-$\omega$ regime (see figure \ref{fig3} (b)). These unique features of $G_{1}(\omega)$ and $G_{2}(\omega)$ indicates towards the dominating effects of negative screening over the RPA (where the full screening is considered). This effect of negative screening introduced by $\lambda(T)$ through $G(\omega)$ in SCR theory is responsible for enhanced correlations in any systems. 

\section{Itinerant magnetism in CaMn$_{2}$Al$_{10}$: DFT studies}\label{sec4}
\begin{figure}[h]
    \centering
\includegraphics[scale = 0.30]{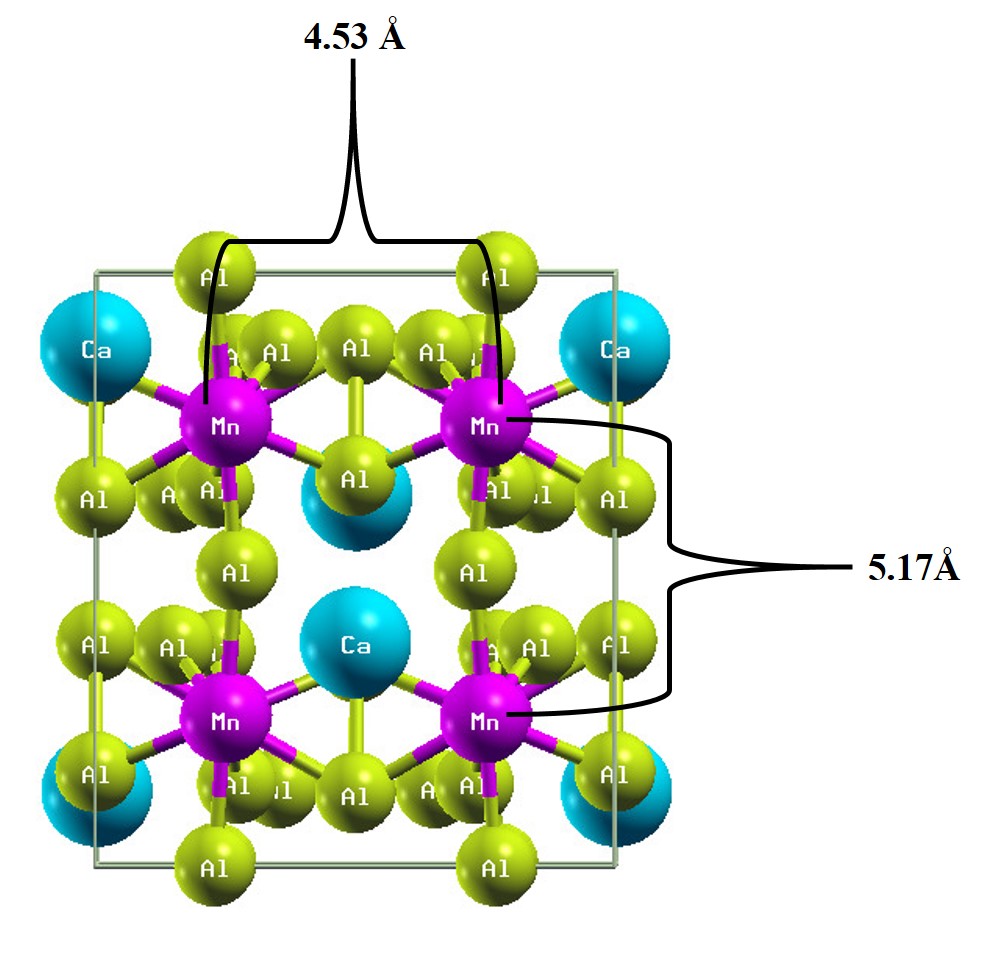}
    \caption{Crystal structure of CaMn$_{2}$Al$_{10}$} 
    \label{f4}
\end{figure}
\begin{figure}
    \centering
    \includegraphics[scale = 0.47]{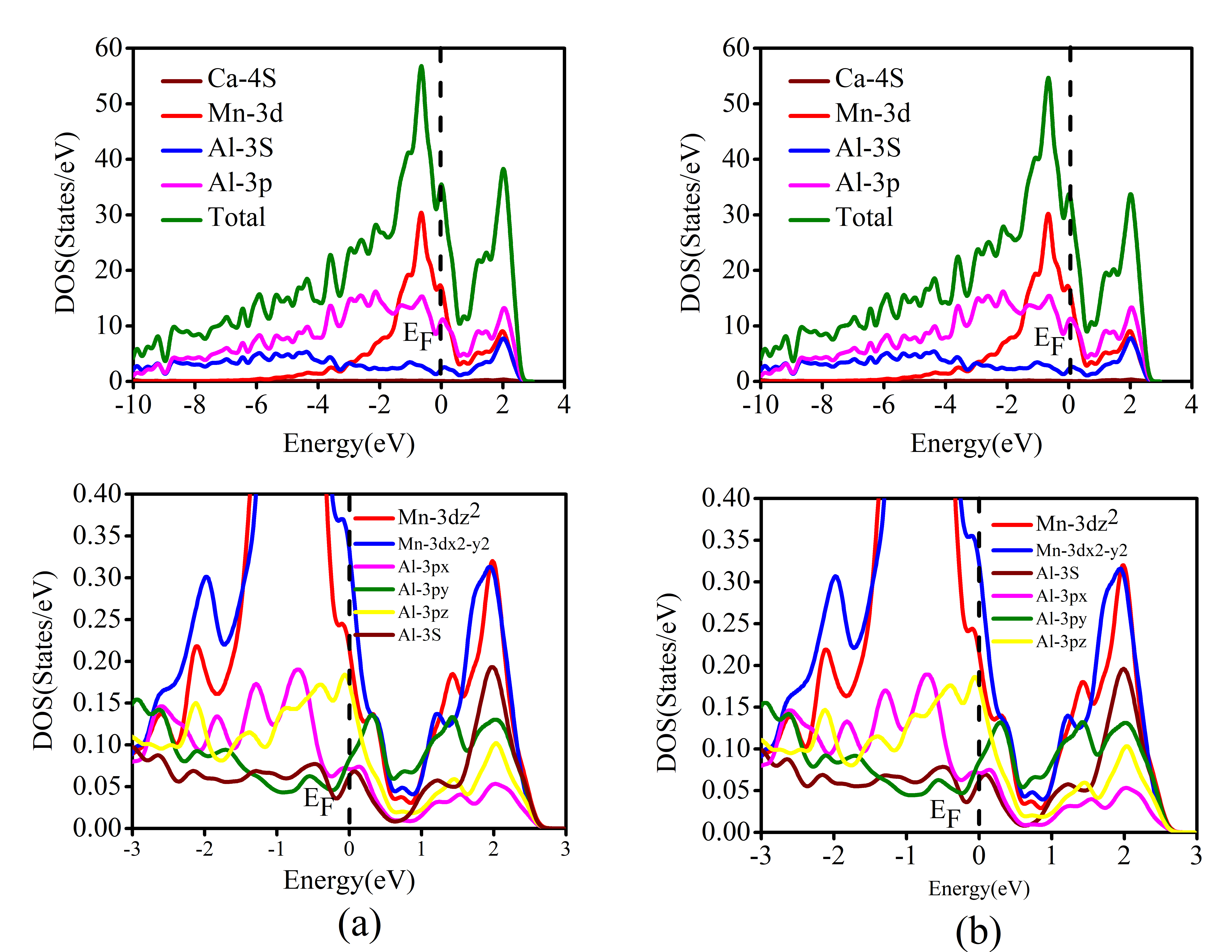}
    \caption{Total DOS and PDOS for CaMn$_{2}$Al$_{10}$ with orbital contributions from (a)DFT calculations and (b) DFT+U calculations}
    \label{f5}
\end{figure}
In Figure \ref{f4}, the highly ordered crystal structure of CaMn$_{2}$Al$_{10}$ is visualized, consisting of two Mn sublattices. The crystal structure is tetragonal with optimised lattice parameter $a = b = 12.6473$ and $12.7165$ Å, c = 4.9986 and 5.0614 Å in DFT and DFT+U calculations respectively\cite{steinke2015camn}. The DFT+U method is a simplified rotationally invariant formalism by the Dudarev et al \cite{dft4}. We have used the value of $U = 0.31$ in the DFT calculations, which is calculated from the SCR theory. This $U$ is applied for the strongly correlated Mn-$3d$ electrons. The lattice parameters calculated with DFT+U are in close agreement with the experimental values \textit{i.e.}, $a = b = 12.8452$ Å and $c = 5.1392$ Å.  All Mn sites are equivalent, with a nearest-neighbour Mn-Mn distance of  $4.53$ Å along the $[110]$-direction, and $5.17$ Å along $[001]$-direction. The total magnetic moment obtained in the spin-polarized DFT (DFT+U) calculations is $8.75(9.69)$ $\mu B/cell$ for the CaMn$_{2}$Al$_{10}$ compound. In DFT (DFT+U), partial magnetic moments of Ca, Mn and Al atoms are about $0.0117( -0.0153)$, $0.934(1.203)$ and $-0.0102(-0.0124)$ $\mu B/cell$ respectively. Spin polarized calculations for the FM state reveal a magnetic moment of $0.93$ $\mu B/cell$ on the Mn site, which is close to the previously reported DFT calculation of $0.9$ $\mu B$ and the observed CW magnetic moment of $0.83$ $\mu B$ \cite{steinke2015camn}. 
 
 Magnetic moment of Mn atoms are anti-parallel to Ca and Al atom, which confirms the ferromagnetic nature of the Compound. Larger electronic density peak in the Mn-DOS near the Fermi level reflects the major contribution of Mn atoms in  magnetization. The orbital contribution of CaMn$_{2}$Al$_{10}$ can be obtained from the partial density of the states (PDOS). We have therefore investigated the PDOS for both DFT and DFT+U calculations.  The total DOS and  partial DOS along with the orbital contributions is given for DFT calculations in fig.\ref{f5}(a) and for DFT+U calculations in fig.\ref{f5}(b). In both the calculations the following signatures of hybridisation between the Mn-$3d$ and Al-$3p$ orbitals is observed:(i) The presence of a pseudogap at the Fermi level indicates hybridisation between the transition metal $d$ and Al $p$ orbitals \cite{hybrid1}. In both fig.\ref{f5}(a) and \ref{f5}(b) a pseudogap is observed suggesting the hybridisation between Mn $3d$ and Al $3p$ orbitals. (ii) In fig.\ref{f5}(a) and \ref{f5}(b) there is an overlap of the Mn PDOS and Al PDOS between $0-1$ eV energy range which indicates hybridisation. It is even clearer in the orbital resolved PDOS where there is overlap of Mn $3d_{z^{2}}$ and Mn $3d_{x^{2} - y^{2}}$ with Al $3 p_{y}$ in the energy range $0$ to $1$ eV. Similarly there is overlap of Mn $3d_{z^{2}}$ and Mn $3d_{x^{2} - y^{2}}$ with Al $3 p_{z}$ near the Fermi energy. Also there is an overlap of Al $3 p_{x}$ and Al $3 p_{z}$ in the energy range $-1$ to $0$ eV. 
 In the partial DOS of DFT+U calculation at the Fermi level Mn-$3d_{x^{2}-y^{2}}$ orbitals contribution (0.38 States/eV) is higher than DFT (0.33 States/eV) calculation. When the Hubbard parameter ($U$) is implemented, Mn-$3_{d}$ states originally lying close to the Fermi level are shifted to higher binding energy due to onsite Coulomb-interaction among Mn-3d electrons. This will close the bands at the Fermi level in DFT+U calculations. Due to this effect the width of the pronounced pseudo gap at the Fermi level in the spin-up bands is slightly increased. 

\section{Conclusion}\label{sec5}
The inverses magnetic susceptibility in the system CaMn$_{2}$Al$_{10}$ obey the Curie-Weiss law as determined from the experiment. We theoretically examined the problem and shows that SCR theory can explain the observed Curie-Weiss law. With this we (1) verify that  CaMn$_{2}$Al$_{10}$ is a case of itinerant magnetism, and  (2) our comparison estimated the electronic correlations in this system , (3) we perform DFT+U calculations (U determined via the SCR theory) and find that it is the strong Mn-Al hybridization that is responsible for itinerant magnetism in this system. Our estimated value of electronic correlations can be potentially used  to address the optical conductivity and many other probes.  \\\\
{\bf{Acknowledgment}}\\\\
This work is supported by Physical Research Laboratory (PRL), Department of Space, Goverment of India. Computations were performed using the HPC resources (Vikram-100 HPC) project at PRL.

\section*{Appendix}
\subsection*{Computation details}
In this section we give some remarks on the computational aspects that should be kept in mind while devising the algorithm for $G(\omega)$ numerical calculations. It is important to note that we are not giving the complete algorithm but only limited to making some comments which are useful in designing the algorithm for computation. 
$G(\omega)$ is the quantity of central importance and its detailed calculations are as follows
$$
f_{0}(q, \omega)=f_{0}^{\prime}(q, \omega)+i f_{0}^{\prime \prime}(q, \omega)
$$
\begin{equation}
\begin{aligned}
G(\omega)=&-\int \mathrm{d} q q^{2}\left[E ( q , \omega ) \left\{\alpha(1+\delta)\left[f_{0}^{\prime \prime}\left(\partial^{2} f_{0}^{\prime} / \partial B^{2}\right)+f_{0}^{\prime}\left(\partial^{2} f_{0}^{\prime \prime} / \partial B^{2}\right)\right]\right.\right.\\
&\left.-\alpha\left[\left(f_{0}^{\prime}\right)^{2}+\left(f_{0}^{\prime \prime}\right)^{2}\right]\left(\partial^{2} f_{0}^{\prime \prime} / \partial B^{2}\right)\right\}
+2[E(q, \omega)]^{2}\left\{\left[\left(\delta+1-f_{0}^{\prime}\right)^{2}-\left(f_{0}^{\prime \prime}\right)^{2}\right]
\left(\partial f_{0}^{\prime} / \partial B\right)\left(\partial f_{0}^{\prime \prime} / \partial B\right)\right.\\
&\left.\left.+\left(\delta+1-f_{0}^{\prime}\right) f_{0}^{\prime \prime}\left[\left(\partial f_{0}^{\prime} / \partial B\right)^{2}-\left(\partial f_{0}^{\prime \prime} / \partial B\right)^{2}\right]\right\}\right]
\end{aligned}
\end{equation}
where
\begin{eqnarray}
E(q, \omega)=\left[\left(\delta+1-f_{0}^{\prime}\right)^{2}+\left(f_{0}^{\prime \prime}\right)^{2}\right]^{-1}
\end{eqnarray}
Important expressions for the real and imaginary parts of the free electron model are as follows
\\\\
Real parts
\begin{equation}
\begin{aligned}
f_{0}^{\prime}(q, \omega)=(1 / 2)-\left\{\left[q^{4}-(4-2 \omega) q^{2}+\omega^{2}\right] / 16 q^{3}\right\} \log \left|\left(q+q_{1}\right)\left(q+q_{2}\right) /\left(q-q_{1}\right)\left(q-q_{2}\right)\right| \\
-\left\{\left[q^{4}-(4+2 \omega) q^{2}+\omega^{2}\right] / 16 q^{3}\right\} \log \left|\left(q+q_{0}\right)\left(q+q_{3}\right) /\left(q-q_{0}\right)\left(q-q_{3}\right)\right|, \\
{\left[\partial f_{0}^{\prime}(q, \omega) / \partial B\right]_{B=0}=\left(\omega / 4 q^{3}\right) \sum_{m=0}^{3} \log \left|\left(q+q_{m}\right) /\left(q-q_{m}\right)\right|,} \\
{\left[\partial^{2} f_{0}(q, \omega) / \partial B^{2}\right]_{B=0}=\left(1 / 2 q^{2}\right)\{\left[q^{4}-(4-\omega) q^{2}+4 \omega\right]\left(q^{2}-q_{1}{ }^{2}\right)^{-1}\left(q^{2}-q_{2}^{2}\right)^{-1}+\left[q^{4}-(4+\omega) q^{2}\right.} \\
\quad-4 \omega]\left(q^{2}-q_{0}^{2}\right)^{-1}\left(q^{2}-q_{3}{ }^{2}\right)^{-1}-q^{-1}\left[1+\left(q^{2} / 4\right)\right] \sum_{m=0}^{3} \log \left|\left(q+q_{m}\right) /\left(q-q_{m}\right)\right| \}
\end{aligned}
\end{equation}
\\
Imaginary parts
\begin{eqnarray}\label{A2}
\begin{aligned}
f_{0}^{\prime \prime}(q, \omega)=(\pi / 4)(\omega / q) \theta\left(q-q_{1}\right) \theta\left(q_{2}-q\right) 
\quad+\left(\pi / 16 q^{3}\right)\left(q^{2}-q_{0}^{2}\right)\left(q_{3}{ }^{2}-q^{2}\right)\\\left[\theta\left(q-q_{0}\right) \theta\left(q_{3}-q\right)-\theta\left(q-q_{1}\right) \theta\left(q_{2}-q\right)\right]
=\left(\pi / 16 q^{3}\right)\left[\left(q^{2}-q_{0}^{2}\right)\left(q_{3}^{2}-q^{2}\right) \theta\left(q-q_{0}\right) \theta\left(q_{3}-q\right)\right. \\
\left.-\left(q^{2}-q_{1}{ }^{2}\right)\left(q_{2}{ }^{2}-q^{2}\right) \theta\left(q-q_{1}\right) \theta\left(q_{2}-q\right)\right], \\
{\left[\partial f_{0}{ }^{\prime \prime}(q, \omega) / \partial B\right]_{B=0}=(\pi / 4)\left(\omega / q^{3}\right)\left[\theta\left(q-q_{0}\right)\left(q_{3}-q\right)-\theta\left(q-q_{1}\right) \theta\left(q_{2}-q\right)\right],} \\
{\left[\partial^{2} f_{0}{ }^{\prime \prime}(q, \omega) / \partial B^{2}\right]_{B=0}=-\left(\pi / 2 q^{3}\right)\left[1+\left(q^{2} / 4\right)\right]\left[\theta\left(q-q_{0}\right) \theta\left(q_{3}-q\right)-\theta\left(q-q_{1}\right) \theta\left(q_{2}-q\right)\right]} \\
\quad+\left(\pi / 8 q^{2}\right)\left\{(1+\omega)^{-1 / 2}\left[q_{0}{ }^{2} \delta\left(q-q_{3}\right)+q_{3}{ }^{2} \delta\left(q+q_{0}\right)\right]\right. 
\left.\quad-(1-\omega)^{-1 / 2}\left[q_{2}{ }^{2} \delta\left(q-q_{1}\right)+q_{1}{ }^{2} \delta\left(q-q_{2}\right)\right]\right\}
\end{aligned}
\end{eqnarray}

where
$$
\left.\begin{array}{l}
q_{1} \\
q_{2}
\end{array}\right\}=1 \mp(1-\omega)^{1 / 2}
$$

$$
\left.\begin{array}{l}
q_{0} \\
q_{3}
\end{array}\right\}=1 \mp(1+\omega)^{1 / 2}
$$

$$
\begin{array}{l}
s=\omega / q \\\\
F(x)=(1 / 2)\left\{1+\left[\left(1-x^{2}\right) / 2 x\right] \log |(1+x) /(1-x)|\right\}
\end{array}
$$
This $G(\omega)$ is the correction to the conventional RPA theory. It is found that $G(\omega)$ constituted of $\alpha \, G_{1}(\omega)$ and $G_{2}(\omega)$. For obtaining  $G(\omega)$ we require both $G_{1}(\omega)$ and $G_{2}(\omega)$ in long wavelength limit. Mathematical concepts involved in the numerical integration of both $G_{1}(\omega)$ and $G_{2}(\omega)$ plays an important role in analytical calculations of spin fluctuations \cite{moriya1965ferro}. The numerical integration of $G_{1}(\omega)$ involves functions like : $f_{0}^{\prime}(q,\omega)$, $f_{0}^{\prime \prime}(q,\omega)$, ${\partial^{2} f_{0}^{\prime}(q,\omega)}/{\partial^{2} B}$ and ${\partial^{2} f_{0}^{\prime \prime}(q,\omega)}/{\partial^{2} B}$.  These functions and their derivatives consists of various $\theta-$ and $\delta-$ functions. The parts of calculations involving the $\delta-$ functions is quite straight forward except for the singularity corrections. It is observed here that the roots of the functions inside the $\theta-$ and $\delta-$ functions plays an important role. In our single band model the $q$-value is restricted up to $q_{c} = 2$, the integration is performed form $q = 0$ to $q = q_{c} =2$. For $\omega > 1$, the contribution comes from only one root which lies inside the region of integration. For $\omega > 8$, there is no contribution from any root in the calculations of $G_{1}(\omega)$. In $\omega < 1$ region,  the singularity correction conditions are used to estimate the contributions of roots of functions inside the $\theta-$ and $\delta-$  functions. Unlike $G_{1}(\omega)$, the numerical integration of $G_{2}(\omega)$ is straightforward as it involves only $\theta-$ functions and does not need any singularity corrections. For $\omega > 1$, the theta function involving the contributions of at least one root upto $\omega = 8$ and there is no contribution thereafter. For $\omega < 1$, it is found that the contribution of all the roots are quiet straightforward and can be calculated by using pre-defined $\theta-$ functions that are readily available which is not true for $G_{1}(\omega)$.
Hence, the calculations $G(\omega)$ are crucial and play an important role in the settling of algorithm for the calculations of other properties \cite{toru}. So far, numerical calculation of $G(\omega)$ has been carried out for the free electron gas model and the results are shown in Figures (\ref{fig2}) and (\ref{fig3}). 
%
here $G_{1}(\omega)$ and $G_{2}(\omega)$ include only $\delta$ as a varying parameter. Numerical results at $q=2$ are shown for entire $\omega$-range. For $\omega>1$,  $G_{1}(\omega)$ becomes negative and reflects the attractive nature of spin correlations in such systems. $G(\omega)$ is the sum of $\alpha \, G_{1}(\omega)$ and $G_{2}(\omega)$. Thus for obtaining  $G(\omega)$ we need to obtain both $G_{1}(\omega)$ and $G_{2}(\omega)$. We will next discuss separately the concepts involved in the numerical integration of both $G_{1}(\omega)$ and $G_{2}(\omega)$ one by one. \\\\
The numerical integration of $G_{1}(\omega)$ is the most involved. The first thing one should note is that for obtaining $G_{1}(\omega)$, we will have to integrate an integrand which is a function of $q$, $f_{0}^{\prime}(q,\omega)$, $f_{0}^{\prime \prime}(q,\omega)$, $\frac{\partial^{2} f_{0}^{\prime}(q,\omega)}{\partial^{2} B} |_{B = 0}$ and $\frac{\partial^{2} f_{0}^{\prime \prime}(q,\omega)}{\partial^{2} B} |_{B = 0}$. Since $\frac{\partial^{2} f_{0}^{\prime \prime}(q,\omega)}{\partial^{2} B} |_{B = 0}$ consists of various theta and delta functions one would need to evaluate the integration with the part involving the theta functions and the delta functions separately and then take the sum. The part involving the delta functions is quite straight forward except for something that we would like to call "Singularity Corrections". Before getting in to the details let us recall the definitions of some special points $q_{0}$,$q_{1}$,$q_{2}$,$q_{3}$,
$$
\left.\begin{array}{l}
q_{1} \\
q_{2}
\end{array}\right\}=1 \mp(1-\omega)^{1 / 2}
$$
$$
\left.\begin{array}{l}
q_{0} \\
q_{3}
\end{array}\right\}=1 \mp(1+\omega)^{1 / 2}
$$
These points are special because they are the roots of the functions inside the theta and delta functions. Since the integration of a delta function gives the functional value of the integrand without the delta function at the point where the delta function peaks provided it is inside the region of integration and is zero otherwise. For the purpose of understanding, if we consider $q_{c} = 2$, the integration is performed form $q = 0$ to $q = q_{c} =2$. When we consider the domain $\omega > 1$, then only $|q_{0}|$ lies inside the region of integration for $1<\omega<8$ and the only contribution comes from this term. For $\omega > 8$, there is no peak inside the region of integration thus the contribution is zero. If we consider the domain $\omega < 1$, then there are three peaks corresponding to $|q_{0}|$, $q_{1}$ and $q_{2}$ inside the region of integration. Note that if we put the values of $|q_{0}|$, $q_{1}$ or $q_{2}$ directly in $f_{0}^{\prime}$ or $\frac{\partial^{2} f_{0}^{\prime}(q,\omega)}{\partial^{2} B} |_{B = 0}$ it diverges. To avoid this one has to add a small value inside the logarithm to obtain a meaningful value and this is exactly what we called the "Singularity Corrections". The contributions from each of these peaks should be calculated separately. Next while integrating the parts involving the theta functions, as in the above case for $\omega > 1$, the only non zero contribution comes from the theta function involving $|q_{0}|$ for $1<\omega<8$ and there are no non zero contributions for $\omega > 8$. It is important to note that $\frac{\partial^{2} f_{0}^{\prime}(q,\omega)}{\partial^{2} B} |_{B = 0}$ quadratically diverges for $|q_{0}|$ and one should use some logical conditions to remove the points where the integrand diverges. For $\omega < 1$, there are non zero contributions from the theta functions involving  $|q_{0}|$, $q_{1}$ and $q_{2}$. Here It is important to note that the function $\frac{\partial^{2} f_{0}^{\prime}(q,\omega)}{\partial^{2} B} |_{B = 0}$ quadratically diverges for $|q_{0}|$, $q_{1}$ and $q_{2}$ and similar to the 
$\omega > 1$ case we will have to remove the points where the functional value of the integrand is not finite. One should also keep in my mind that since the integrand drastically peaks near these points, one will have to take more points near these regions for integration to get a satisfactory accuracy up on numerical integration. 
\\\\
{\bf Numerical integration of $G_{2}(\omega)$}\\\\
The numerical integration of $G_{2}(\omega)$ is straightforward when compared to $G_{1}(\omega)$, as it involves only theta functions and does not need any "Singularity corrections". As before for $\omega > 1$ only the theta function involving $|q_{0}|$ contributes until $\omega = 8$ and there is no contribution thereafter. For $\omega < 1$, there are contributions from $|q_{0}|$, $q_{1}$ and $q_{2}$ and can be calculated quiet straightforwardly either by designing a code for the theta function or using pre-defined theta functions that are readily available.\\\\
{\bf Details of DFT calculations }\\\\
First principles calculations of structural, electronic and magnetic properties were done using the projector augmented-wave (PAW) method implemented within the QUANTUM ESPRESSO simulation package \cite{dft1}. Generalized Gradient approximation GGA) was used with Perdew-Burke-Ernzerhof (PBE) exchange-correlation functional \cite{dft2}. The structural optimization was done with the gamma-centred Monkhrost-Pack of $12×12×12$ k-point mesh within an energy cut-off of $60$Ry. All the atoms and cell parameters are relaxed by minimizing the inter-atomic force up to $5$ meV/Å. The Ca (4s), Mn (3d) and Al (3s-3p) are treated as a valence states. The on-site Coulomb interaction parameter $U$ and exchange parameter $J$ (here $J=0$) are combined into a single parameter $U_{eff} = U – J$.
\bibliographystyle{unsrt}
\bibliography{bib}
\end{document}